\documentclass[prb,preprint,preprintnumbers,amsmath,amssymb]{revtex4-1}

\usepackage{graphicx}
\usepackage{dcolumn}
\usepackage{bm}
\usepackage{color}

\begin{document}


\def\Ef{$E_{\rm F}$}
\def\Ed{$E_{\rm D}$}
\def\Eg{$E_{\rm g}$}
\def\Efmath{E_{\rm F}}
\def\Edmath{E_{\rm D}}
\def\Egmath{E_{\rm g}}
\def\Tc{$T_{\rm C}$}
\def\kpara{{\bf k}$_\parallel$}
\def\kparamath{{\bf k}_\parallel}
\def\minuskpara{$-{\bf k}_\parallel$}
\def\kparazero{{\bf k}$_{\parallel,0}$}
\def\minuskparazero{$-{\bf k}_\parallel,0$}
\def\kperp{{\bf k}$_\perp$}
\def\invA{\AA$^{-1}$}
\def\Kbar{$\overline{\rm K}$}
\def\Gbar{$\overline{\Gamma}$}
\def\Mbar{$\overline{\rm M}$}
\def\GbarMbar{$\overline{\Gamma}$-$\overline{\rm M}$}
\def\GbarKbar{$\overline{\Gamma}$-$\overline{\rm K}$}
\def\DeltaEf{$\Delta E_{\rm F}$}
\def\DeltaEfmath{\Delta E_{\rm F}}
\def\BiTe{Bi$_2$Te$_3$}
\def\BiSe{Bi$_2$Se$_3$}
\def\BiSnTe{(Bi0.67\%Sn)$_2$Te$_3$}
\def\SbTe{Sb$_2$Te$_3$}
\def\MS{\textcolor{red}{}}

\title{ Tolerance of   topological surface states towards magnetic moments: Fe on \BiTe\ and \BiSe}
  \author{ M. R. Scholz,$^1$ J. S\'anchez-Barriga$^1$, D. Marchenko,$^1$ A. 
Varykhalov,$^1$    
           A. Volykhov,$^2$ L. V. Yashina,$^2$ and O. Rader$^1$ }
 \address{$^1$Helmholtz-Zentrum Berlin f\"ur Materialien und Energie, 
Elektronenspeicherring BESSY II, Albert-Einstein-Str. 15, D-12489 Berlin, Germany}
 \address{$^2$Department of Chemistry, Moscow State University, 
          Leninskie Gory, 1/3, 119992 Moscow, Russia}
  
\begin{abstract}

{\bf 

Topological 
insulators \cite{KanePRL2005,FuPRB2007,BernevigPRL06,FuPRL2007,MoorePRB07,Roy06,Murakami,KoenigScience2007}
are a novel form of matter which features metallic surface states
with quasirelativistic dispersion similar to graphene \cite{RotenbergNP11}.
Unlike graphene, the locking of spin and momentum and the protection by time-reversal
symmetry\cite{KanePRL2005,FuPRB2007,BernevigPRL06,FuPRL2007,MoorePRB07,Roy06,Murakami,KoenigScience2007}
open up tremendous additional possibilities for external control of transport 
properties\cite{QiNP08,QiPRB08,YokoyamaPRB08,GaratePRL10,MondalPRL10,WuPRB10,YZhangAPL10,BurkovPRL10,KongAPL11}.
Here we show by angle-resolved photoelectron spectroscopy that the topological surface states 
of \BiTe\ and \BiSe\ are stable against the deposition of Fe without opening a band gap.
This stability extends to low submonolayer coverages meaning that the band gap 
reported recently\cite{WrayNatPhys2010} for Fe on \BiSe\ is incorrect as well as to complete monolayers meaning
that  topological surface states   can very well  exist at interfaces with ferromagnets
in future devices.  

      } 

\end{abstract}


\maketitle

In contrast to most ordinary insulators, the bulk band gap of a topological insulator is
caused by band inversion and strong spin-orbit coupling. Its surface is metallic due to 
nondegenerate and spin polarized surface states 
\cite{FuPRL2007,MoorePRB07,Roy06,Murakami}.
The spin is such that two surface-state electrons that propagate in opposite
directions carry opposite spins. 
Owing to this spin chirality,  
electron backscattering by 180$^\circ$ is forbidden as it would require a spin flip.
The surface state itself enjoys   protection  by time-reversal symmetry against distortions such
as surface impurities  \cite{FuPRL2007,MoorePRB07,Roy06,Murakami}.

On the other hand, a magnetic field breaks time-reversal symmetry and by lifting the degeneracy
$E(\uparrow) = E(\downarrow)$ causes a gap in the topological surface state at zero momentum
in inversion-symmetric systems.  
 This has been proven in transport measurements by applying 28 mT to a HgTe-based 
two-dimensional quantum spin Hall system  \cite{KoenigScience2007}. 
For a three-dimensional topological insulator, 
much more efficient than an external field is the use of the exchange field from a 
deposited ferromagnet. 
For such configuration, numerous  electronic effects   have been
 predicted\cite{QiNP08,QiPRB08,YokoyamaPRB08,GaratePRL10,MondalPRL10,WuPRB10,YZhangAPL10,BurkovPRL10,KongAPL11}
such as the quantized topological magnetoelectric effect leading to half-integer charges
at magnetic domain boundaries \cite{QiPRB08,QiNP08}
a magnetoresistance increasing for parallel (sic!) domain magnetizations \cite{YokoyamaPRB08}, 
and an inverse spin-galvanic effect enabling the dissipationless  magnetization reversal
by an applied electrical current \cite{GaratePRL10}. 

While these effects are awaiting experimental verification, much progress has recently been 
made in the identification of appropriate topological three-dimensional insulator materials,
starting with Bi10\%Sb \cite{HsiehNature08}. 
The second generation of three-dimensional topological insulators, \BiSe\ and 
\BiTe\ \cite{FuPRB2007,ZhangNatPhys2009,HsiehNature2009} forms 
particularly simple systems with the topological 
surface state appearing as a single Dirac cone at the \Gbar-point of the surface
Brillouin zone, i. e., for $k_\parallel=0$, and a substantial spin polarization 
\cite{HsiehNature2009} which has been shown to approach 100\%\ \cite{Scholz10,PanPRL10}.
The band gaps are large enough to allow for room-temperature applications 
but as-grown samples show a tendency for defects (e. g., excess Te on Bi sites)
which lead to a $n$-type bulk doping  so large that the samples are rendered 
metallic in their bulk.  

Several studies have investigated the effect of magnetic moments on the topological surface state. 
When \BiSe\ is doped with Fe in the bulk, a gap opens in the surface-state dispersion at  
Fe concentrations (relative to Bi) from 5\%\ to 25\%\  \cite{ChenScience2010}. 
This includes concentrations for which the system remains 
paramagnetic down to low temperatures ($T = 2$ K), such as 12\%\ Fe, where the 
 surface-state band gap amounts to 45 meV \cite{ChenScience2010}. 
When \BiSe\ is p-doped in the bulk by Mn, it is possible to move 
the bulk band gap to the Fermi level \cite{ChenScience2010}. 
Bulk Mn doping of 
\BiTe\ and \SbTe\ has led to ferromagnetic order  below temperatures of 10 K and 17 K, respectively. 
   Unfortunately, photoemission of the Dirac point
region remained inconclusive, possibly due to the strong p-doping \cite{HorPRB2010}.

Most recently, Fe deposited directly on the 
surface of \BiSe\ has been studied  by means of angle-resolved photoemission \cite{WrayNatPhys2010}. 
It was revealed that the Fe  dopes the surface substantially beyond its self-doping and opens 
at submonolayer coverages a surface-state band gap of 100 meV. The authors speculate
that the topological surface state mediates \cite{LiuPRL2009} an out-of-plane anisotropy of the 
Fe magnetic moments without long-range ferromagnetic order \cite{WrayNatPhys2010}. 

In the present Letter, we investigate the behavior of \BiTe\ and \BiSe\ upon Fe deposition. 
We characterize the system Fe/\BiTe\ systematically by angle-resolved photoemission
of the valence band and by photoemission of the Bi 5d, Te 4d, and Fe 3p core levels.
In agreement with previous work \cite{WrayNatPhys2010} we observe energy shifts of the
surface band structure consistent with electron doping. 
We find that we can shift the Dirac point for both systems by not more than 150 meV which is
much less than what was reported previously. 
Most strikingly we show direct evidence that the Dirac crossing remains intact even under 
heavy Fe deposition and that in no stage of Fe deposition a band gap opens.

For \BiTe\  we monitor the evolution of the surface state band structure with 55 eV photons at room 
temperature after each deposition step of Fe. Panel a of Figure 1 shows the $E({\bf k})$ dispersion 
relations of the topological surface state 
along the \Mbar-\Gbar-\Mbar\ direction of the surface Brillouin zone (see inset)  
 with a red dashed line showing the shift of the Dirac point with increasing Fe deposition. 
In the pristine sample (top left, 0 ML) we see the   V-shaped dispersion of the topological surface
state like in previous work \cite{ChenScience2009}. 
The Dirac crossing 
point   is clearly visible at the present photon energy where the intensities of the 
two branches appear to add up. 
The Dirac point lies in a dip of the bulk valence band projection at 200 meV binding energy 
which has its origin in the band inversion \cite{ZhangNatPhys2009}.
Furthermore, the surface bands hybridize for symmetry reasons \cite{Scholz10} 
with the bulk bands at \Gbar\   which follow an upward dispersion 
but appear weak like in previous work \cite{ChenScience2009,HsiehNature2009}. 

Figure 1 shows for 0.1 ML Fe an 
increasing background and the appearance of a well defined Fermi edge due to Fe 3d states.
Surface bands have shifted rigidly towards higher binding energies by 96 meV. 
 At the center of the surface Brillouin zone the bottom of a new, parabolic shaped band appears 
at 20 meV binding energy. By increasing the amount of 
Fe the band structure shifts further to higher binding energies  
with the deposited mass following an exponential-like behaviour (Fig. 2a). 
Around 0.3 ML Fe  we achieve  
saturation of the surface doping. The shift of the Dirac point reaches 150 meV and
can also be followed for normal-electron-emission (\Gbar) spectra 
in the inset of Fig. 2a. (The appearance of a shoulder
 at the high-binding-energy side of the Dirac point will be discussed further below in connection with
 our low-temperature measurements.)\\
 
  Figure 1b shows the effect of Fe deposition on the topological surface state of \BiSe.
 18 eV photons are used which enhances   the contrast between bulk and  surface states as 
 compared to 55 eV.  Again, the band structure undergoes 
 a shift consistent with n-type surface doping.
The region of the bulk 
 conduction band, situated in between the two branches of the topological surface state, 
alters in a peculiar way. 
At 0.1 ML Fe the Dirac point has shifted by 100 meV, considerably less than the $>300$ meV
shift assigned previously to this coverage\cite{WrayNatPhys2010}. More importantly, 
the shift saturates at 140 meV for 0.3 ML.

As can be seen in Figure 2b the Bi 5d and Te 4d core levels exhibit a similar shift as the Dirac point.
However, while the Te 4d states shift much stronger and the shift does not seem to saturate
 even at 1 ML Fe coverage, the Bi 5d states shift even less than the Dirac point but saturate
 at a similar coverage.
The black curve in the upper inset shows  the spin-orbit split 5d$_{3/2}$ and 5d$_{5/2}$ 
core-level peaks at 27.6 eV and 24.55 eV binding energies,  respectively. 
(For better comparability,  a Shirley background  has been subtracted and the spectra
have been normalized to the same area.)

In addition to the shift to higher binding energies, shoulders 
become with increasing Fe amount visible at the lower binding energy side of each Bi 5d peak (see also inset). 
The stoichiometry of \BiTe\ demands the Bi atoms to be in 
the $3+$ oxidation state ($2-$ for Te) so that the two shoulders are assigned to Bi with lower oxidation state than $3+$. 
Since the topmost layer of the (00$\cdot$1) cleavage plane is made up of Te atoms
\cite{Wyckoff1964}, the Fe is expected to react with Te. 
The newly formed bonds between Fe and Te lead to a weakening of the interlayer bonding between
  Te and Bi atoms of the topmost two atomic layers of the substrate.
 This is followed by a strengthening of the intra-layer
 bonding between Bi atoms of the second layer which causes the chemical shift.
To confirm that the observed changes are really due to Fe, we show the growing Fe 3p core level  
in Fig. 2c for different coverages.
The Bi 5d states in \BiSe\ show the same strong changes as can be seen in Fig. 2d.   
 This leads to the
same microscopic picture as compared to \BiTe. It should be emphasized that Fig. 2d reports a 
fairly strong chemical interaction of Fe with the substrate against which the surface state 
of the substrate remains
topologically protected. This insensitivity is worthwhile mentioning in addition to those towards
 disorder, Coulomb and magnetic perturbations.\\
The analysis of the Se 3d peak is not straightforward since the Fe 3p resides approximately at the 
same binding energy and in addition to growth of this peak underneath we expect changes due to the 
chemical reaction similar as in the Te 4d states. There are indeed similarities like an energy shift
 and an asymmetry at the higher binding energy side. The difference is a change in the relative 
intensities of the Se 3d$_{3/2}$ and the Se 3d$_{5/2}$ peak. We normalized to equal  Se 3d$_{5/2}$ intensity 
to emphasize that the Se 3d$_{3/2}$ peak contains the additional Fe 3p spectral weight.\\



To investigate the effect of the Fe adatoms on the surface electronic structure in more 
detail and to enable the suggested out-of-plane alignment of Fe moments induced at low temperatures
by the topological surface state\cite{WrayNatPhys2010}, 
we have cooled the Fe-covered sample down to $T=50$ K for \BiTe\ and 8 K for \BiSe.  
Upon cooling, the Dirac point has been found to shift to larger binding energies between 50 and 100 meV.
This effect is visible in the core-level spectra and reverses after warming
up. This may be due to temporary residual gas adsorption. 
It is clearly seen in Fig. 3a that the Dirac point remains intact. The 
second derivative data of 0.4 ML Fe shows  in addition that the Dirac point as the origin of the
linearly dispersing  topological surface state  stays  connected to the bulk band.
In the second derivative, also an M-shaped dispersive band is resolved  which is separated from
the Dirac point at \Gbar\ by 100 to 200 meV.  

 In \BiSe\  we again observe a similar picture   (Fig. 3b). We are able to resolve 
 a pair of M-shaped bands in the second derivative which are well separated from the Dirac point. 
No gap has opened at the Dirac point which is in clear contradiction to the previous
 observation\cite{WrayNatPhys2010}.\\
 
  Finally we want to turn to the extra states close to the Fermi level. 
In Fig 3d we show the Fermi surface of \BiTe\ measured with 21 eV photons at low temperature. 
At this energy 
extra states which cannot be seen well at 55 eV due to unfavourable photoemission cross section
 are seen to appear in the cut along the \Kbar-\Gbar-\Kbar\ direction.
We find two nested parabolas with the bottom of the lower band at 60 meV binding energy. 
We also observe a pair of nested parabolas in \BiSe. 
In the previous work, a stronger doping-induced energy shift was observed and a pronounced
$\kparamath$ splitting of Rashba type was resolved  \cite{WrayNatPhys2010}. 
Our doping induced shifts saturate instead at 140 meV and no Rashba-type spin-orbit 
splitting is seen in the well-resolved bands. 
In a most recent work, extra states at the Fermi energy have been obtained instead of 
volutary deposition by just exposing the surface of \BiSe\ to residual gas\cite{BiancchiNatCom2010}.  
Their formation has been explained as a two-dimensional electron gas due to a quantum well  
at the surface of \BiSe\ due to band bending \cite{BiancchiNatCom2010}. 
For the present experiment this means that it is especially important not to cross 
 contaminate the surface by other species during the Fe deposition experiments.

In conclusion, we find that the topological surface state of \BiTe\ is tolerant against
magnetic adsorbates not only at low coverages   but at least up to the 
limits of detectability by angle resolved photoemission which lie around 1 ML for \BiTe\ in the
present experiment. This means that \BiTe\ can be interfaced with a ferromagnet without
losing the topological surface state and its unique dispersion. For \BiSe\ we achieve the same conclusion
thus contradicting the gap opening reported before. We show that Fe reacts chemically when deposited
on the surface which gives additional emphasis to the unique robustness of topological surface states.
The presently found robustness is the precondition for the exploration and the successful 
functionalization of interfaces between topological insulators and ferromagnets.
This will involve growing a perpendicularly magnetized ferromagnetic film on top of a topological
insulator and monitoring the effect of the exchange coupling on the topological surface state underneath.\\ 

{\bf Methods}

Single crystals of \BiTe\ and \BiTe\ were grown by the Bridgman method. Small portions of the  
crystal have been cut and prepared with copper adhesive tape for \textit{in situ} cleavage 
resulting in (00$\cdot$1) surfaces in hexagonal coordinates. Angle-resolved photoemission
 measurements have 
been carried out in ultra high vacuum of $1\cdot10^{-10}$ mbar 
with a Scienta R8000 electron analyzer at the 
UE112-PGM2a beam-line of BESSY II with p polarized undulator radiation. Fe was 
deposited \textit{in situ} from an e-beam oven with the sample kept at room temperature. The evaporation 
rate has been repeatedly calibrated with a quartz microbalance and was typically 0.05 ML/min.

\newpage

\noindent FIG. 1. {\bf Behaviour of topological surface states upon deposition of Fe.}
Angle-resolved photoemission data at room temperature for (a) \BiTe\ and (b) \BiSe. 
The red dashed line follows the surface doping effect on the Dirac point. The 
Dirac point is
clearly observed at the employed photon energies of (a) 55 eV and (b) 18 eV. 

\phantom{Zeile}

\noindent FIG. 2. {\bf Substantial doping and chemical interaction from core-level spectroscopy.}
(a) The surface doping of \BiTe\ from normal-emission valence-band spectra (inset) is roughly 
in agreement with the core-level shifts of (b) Bi 5d and Te 4d, except for Fe 3p 
of the dopant (c). (d) The shoulders of Bi 5d reflect a substantial chemical interaction.
(e) The Fe 3p signal is hidden under Se 3d in \BiSe.

\phantom{Zeile}

\noindent FIG. 3. {\bf The topological surface state at low temperature.}
Fe deposition at room temperature does not open a gap in the topological
surface state on (a) \BiTe\ at 50 K and (c) \BiSe\ at 8 K. 
The Fe-induced states at the Fermi energy appear as a pair in (c) \BiSe\
and in (b) \BiTe\ when measured at low photon energies (21 eV) where also 
their strong warping is revealed. 
Also the valence band becomes split as revealed by two M-shaped dispersions.
This does not affect the Dirac crossing point of the topological surface state. 

\begin{figure}[ht]
	\centering
  \includegraphics[width=1\textwidth]{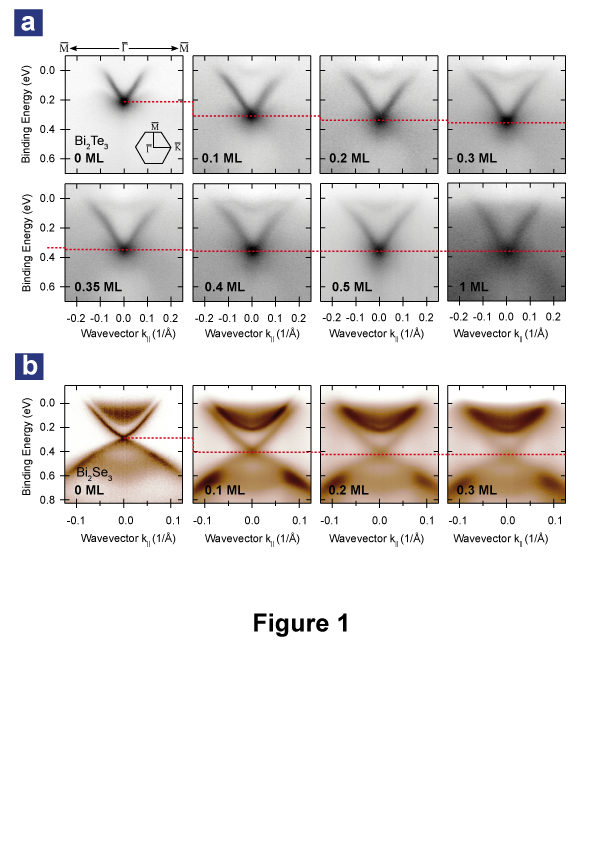}
\end{figure}

\begin{figure}[ht]
	\centering
  \includegraphics[width=1.4\textwidth,angle=90]{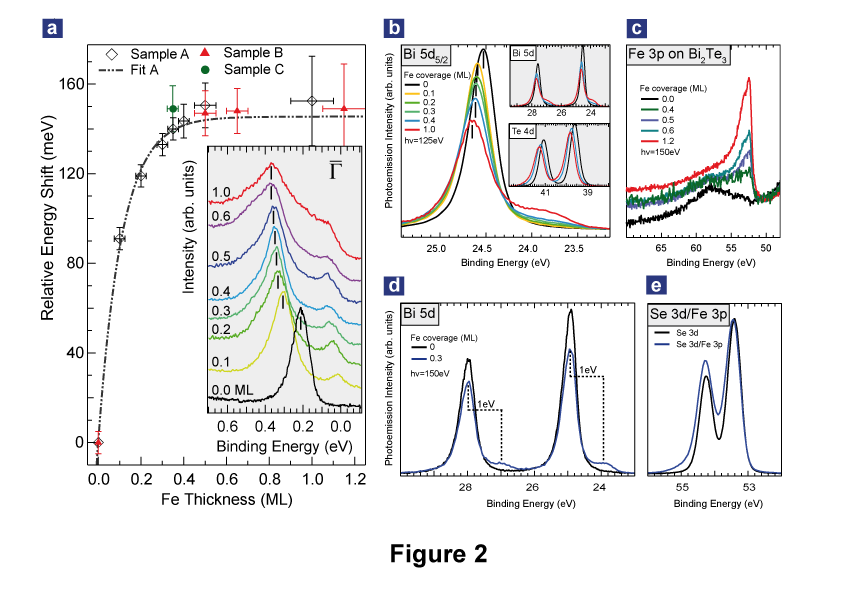}
\end{figure}

\begin{figure}[ht]
	\centering
  \includegraphics[width=1.3\textwidth,angle=90]{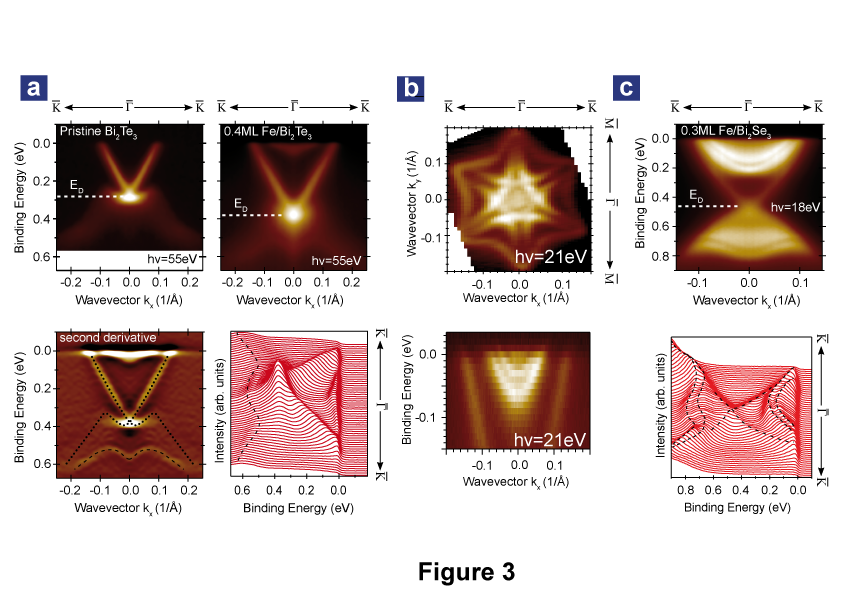}
\end{figure}

\end{document}